\documentclass[aps,twocolumn]{revtex4-1}
\usepackage{amsfonts}
\usepackage{amssymb}
\usepackage{amsmath}
\usepackage{graphicx}
\usepackage{epsfig}
\usepackage{subfigure}
\usepackage{appendix}
\usepackage{color}
\usepackage{hyperref}

\hypersetup{hypertex=true,
colorlinks=true,
linkcolor=blue,
anchorcolor=blue,
citecolor=blue}

\begin{document}

\title{Topological charge pumping in dimerized Kitaev chains}
\author{E. S. Ma}
\author{Z. Song}
\email{songtc@nankai.edu.cn}
\affiliation{School of Physics, Nankai University, Tianjin 300071, China}
\begin{abstract}
We investigated the topological pumping charge of a dimerized Kitaev chain
with spatially modulated chemical potential, which hosts nodal loops in
parameter space and violates particle number conservation. In the simplest
case, with alternatively assigned hopping and pairing terms, we show that
the model can be mapped into the Rice-Mele model by a partial particle-hole
transformation and subsequently supports topological charge pumping as a
demonstration of the Chern number for the ground state. Beyond this special
case, analytic analysis shows that the nodal loops are conic curves.
Numerical simulation of a finite-size chain indicates that the pumping
charge is zero for a quasiadiabatic loop within the nodal loop and is $\pm 1$
for a quasiadiabatic passage enclosing the nodal loop. Our findings unveil a
hidden topology in a class of Kitaev chains.
\end{abstract}

\maketitle

\section*{Introduction}

Thouless pumping \cite{Thouless1983Quan} has received much attention over a
long period of time. It is the quantum version of matter pumping by a
mechanical device in our everyday life. The intriguing features are that (i)
the total probability of the transferred particles is precisely quantized
for a cyclic adiabatic passage without a bias voltage and (ii) a nonzero
pumping charge for the ground state is shown to relate a degenerate point
and then acts as a topological invariant \cite{XiaoDi}.

Recent enhanced quantum manipulation techniques allowing precise control of
the time-dependent periodic potential have made experimental realization of
quantum pumping possible. Electron pumping experiments have been performed
in various semiconductor-based nanoscale devices \cite%
{Switkes1999a,Blumenthal2007,kaestner2008single}. More recently, the
topological charge pump was realized in optical superlattices based on
ultracold atom technology \cite%
{nakajima2016topological,lohse2016thouless,lu2016geometrical}, and it was
also extensively studied in theory \cite%
{chiang1998quantum,qian2011quantum,wang2013topological,matsuda2014topological,mei2014topological,wei2015anomalous,yang2018continuously,Matsuda2022,zhang2020top}%
. To date, both experimental and theoretical studies have focused mainly on
systems that involve the conversion of particles, such as 1D or 2D
topological insulators.

In this work, we studied the topological pumping charge of a dimerized
Kitaev chain model with spatially modulated chemical potential. In general,
the topological pumping charge in a one-dimensional Kitaev model refers to
the corresponding Majorana lattice rather than the transport of spinless
fermions \cite{XiaoDi,WR1,WR2}. Technically, the origin of this topological
feature is the degenerate point. In contrast, the present dimerized Kitaev
model hosts nodal loops in parameter space, and the pumping charge is
obtained directly from the ground state of the model, which does not support
particle number conservation. The main motivation of this work arises from
the simplest case in which the hopping and pairing terms are completely
assigned to different dimers alternatively. We show that this model can be
mapped into a Rice-Mele (RM) model \cite{Rice} by a partial particle-hole
transformation under certain constraints. Then, the ground state of such a
Kitaev model supports the topological charge pumping as a demonstration of
the Chern number because the current operator is invariant under the
transformation. Beyond this special case, analytic analysis has shown that
the degenerate points become nodal loops. Numerical simulation of a
finite-size chain indicates that the pumping charge is zero for a
quasiadiabatic loop within the nodal loop and $\pm 1$ for the passage loop
enclosing the nodal loop. Our findings unveil a hidden topology in a class
of Kitaev chains.

\begin{figure*}[tbh]
\centering
\includegraphics[width=0.9\textwidth]{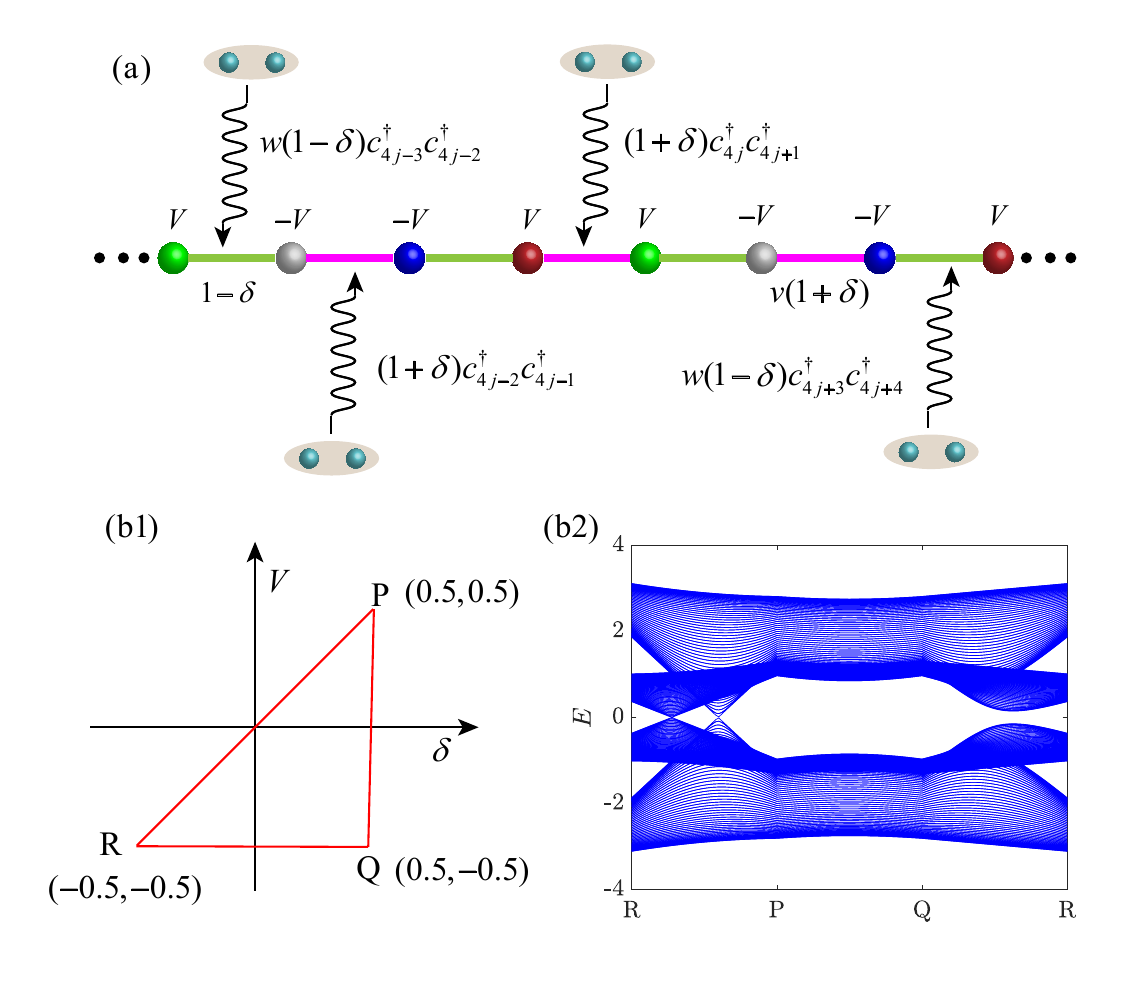}
\caption{(a) A schematic diagram of the Hamiltonian in Eq. (\protect\ref{Ha}%
). A unit cell is composed of four sites with modulated chemical potential.
The strengths of the hopping terms and pairing terms are stagger. $1-\protect%
\delta$ and $v\left( 1+\protect\delta \right)$ are the strengths of the
hopping terms, and $w\left( 1-\protect\delta \right)$ and $1+\protect\delta$
are the strengths of pair creation (annihilation). (b1) shows a triangular
loop on the $\protect\delta -V$ plane, and (b2) shows the corresponding
energy spectra along the loop. The spectra are symmetrical with respect to
zero, and there are two degenerate points because the loop has two crossing
points with the degenerate line of the energy spectra. The parameters are $%
N=50,w=0.6,$ and $v=-0.3$. }
\label{fig1}
\end{figure*}

The rest of the paper is organized as follows. We begin Sec. \ref{Model and
nodal ellipses} by introducing the Hamiltonian and its symmetry. We also
analytically deduce the equations of nodal lines in parameter space. In Sec. %
\ref{Equivanlent RM model} we demonstrate that the model under some
constraints can be mapped into the Rice-Mele model via a partial
particle-hole transformation, which inspires us to investigate the pumping
charge of this model. In Sec. \ref{Topology of the nodal ellipse}, we
numerically calculate the pumping charges of the ground state for different
quasiadiabatic passages in parameter space. The results indicate that the
pumping charge can be treated as a topological invariant to characterize the
topology of the model. Finally, we draw conclusions in Sec. \ref{summary}.
Some detailed derivations are given in the Appendix.

\begin{figure*}[tbh]
\centering \includegraphics[width=0.9\textwidth]{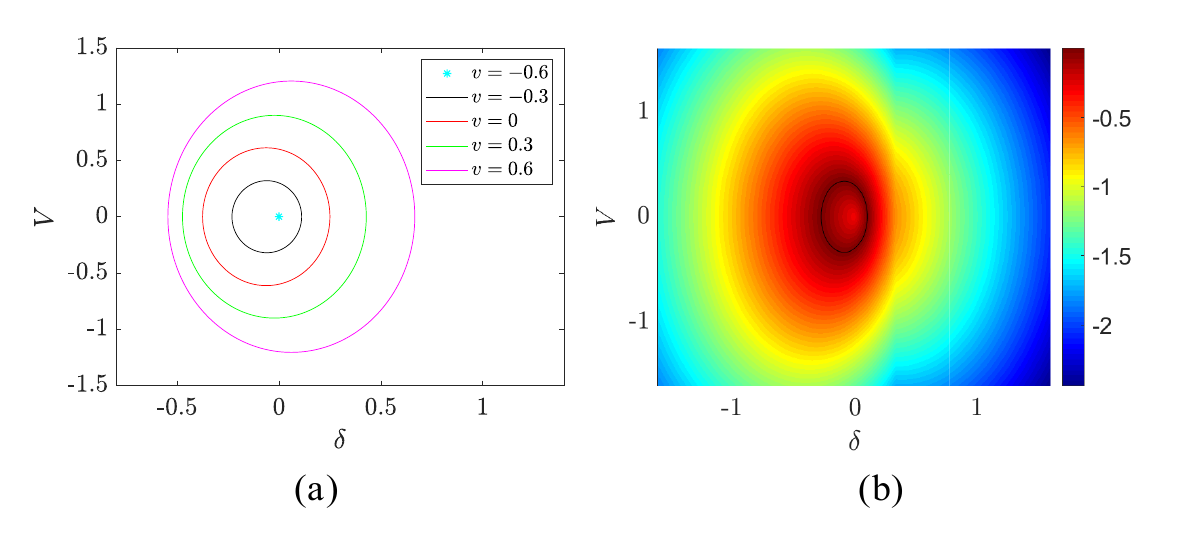}
\caption{(a) Several representative nodal lines derived from Eq. (\protect
\ref{ellipse}) on the $\protect\delta-V$ plane for fixed $w$ and different $%
v $. (b) Color contour plots of the numerical results of one of the energy
bands with the wavevector $k=0$. The black loop is the nodal line from (a)
with $v=-0.3$ and this indicates that the degenerate line on the $\protect%
\delta -V$ plane is a closed loop, which accords with that in (a). The other
parameters are $N=50$ and $w=0.6$. }
\label{fig2}
\end{figure*}

\section{Model and nodal ellipses}

\label{Model and nodal ellipses}

The Kitaev chain model describes a spin-polarized $p$-wave superconductor in
a one-dimensional system and has received much attention since its simple
form also includes a rich phase diagram. This system has a topological
phase, realizing Majorana zero modes at the ends of the chain \cite{Kitaev}.
On the other hand, it is the fermionized version of the well-known
one-dimensional transverse-field Ising model \cite{Pfeuty}, which is one of
the simplest solvable models that exhibit quantum criticality and phase
transition with spontaneous symmetry breaking \cite{SachdevBook}. Several
studies have been conducted with a focus on long-range Kitaev chains, in
which the superconducting pairing term decays with distance as a power law 
\cite{TPCHOY,AGG,DV1,DV2,OV,LL,UB}.

In this work, we investigate the topological features of the Kitaev chain
from an alternative perspective, referred to as the hidden topology. Such a
feature does not emerge in the usual Kitaev chain considered in the
literature. Considering a dimerized one-dimensional Kitaev model, the
Hamiltonian contains two parts 
\begin{equation}
H=H_{1}+H_{2},  \label{Ha}
\end{equation}%
with 
\begin{eqnarray}
H_{1} &=&\sum_{l=1}^{2N}\left[ \left( 1-\delta \right) c_{2l-1}^{\dag
}c_{2l}+\left( 1+\delta \right) c_{2l}^{\dag }c_{2l+1}^{\dag }\right] +%
\mathrm{h.c.}  \notag \\
&&+V\sum_{l=1}^{2N}\left( -1\right) ^{l+1}\left( c_{2l-1}^{\dag
}c_{2l-1}-c_{2l}^{\dag }c_{2l}\right) ,
\end{eqnarray}
and 
\begin{equation}
H_{2}=\sum_{l=1}^{2N}\left[ \left( 1-\delta \right) wc_{2l-1}^{\dag
}c_{2l}^{\dag }+\left( 1+\delta \right) vc_{2l}^{\dag }c_{2l+1}\right] +%
\mathrm{h.c.}.
\end{equation}%
The dimerized Kitaev model is schematically illustrated in Fig. \ref{fig1}
(a). This seems to be somewhat complicated, and the purpose of this
separation is to provide a better presentation for the following
discussions. Here, $c_{j}$\ denotes the fermion annihilation operator at
site $j$. $\left( 1\pm \delta \right) $ are the hopping amplitudes and the
strength of the pairing operator between neighboring sites with dimerized
factors $1$, $v$, and $w$, and the real number $V$ is the chemical
potential. When the periodic boundary is taken, we define $c_{4N+j}=c_{j}$.

For arbitrary parameters, the unit cell includes four sites. Then, the
Hamiltonian $H$ can be written in block diagonal form%
\begin{equation}
H=\sum_{\pi >k>0}\Psi _{k}h_{k}\Psi _{k}^{\dag }=\sum_{\pi >k>0}\Psi
_{k}\left( 
\begin{array}{cc}
A & -B \\ 
B & -A%
\end{array}%
\right) \Psi _{k}^{\dag },
\end{equation}%
by introducing Fourier transformation%
\begin{equation}
\left( 
\begin{array}{c}
c_{4j-3} \\ 
c_{4j-2} \\ 
c_{4j-1} \\ 
c_{4j}%
\end{array}%
\right) =\frac{e^{ikj}}{\sqrt{N}}\sum\limits_{k}\left( 
\begin{array}{c}
\alpha _{k} \\ 
\gamma _{k} \\ 
\beta _{k} \\ 
\eta _{k}%
\end{array}%
\right) ,
\end{equation}%
with $k=2\pi m/N$, $m=1,2,\cdots N$. Here, the operator vector is defined as 
\begin{equation}
\Psi _{k}=\left( 
\begin{array}{cccccccc}
\alpha _{k}^{\dag } & \gamma _{k}^{\dag } & \beta _{k}^{\dag } & \eta
_{k}^{\dag } & \alpha _{-k} & \gamma _{-k} & \beta _{-k} & \eta _{-k}%
\end{array}%
\right) ,
\end{equation}%
and the two $4\times 4$ matrices $A$\ and $B$\ are

\begin{equation}
A=\left( 
\begin{array}{cccc}
V & 1-\delta & 0 & v\Gamma _{-k} \\ 
1-\delta & -V & v\Gamma _{0} & 0 \\ 
0 & v\Gamma _{0} & -V & 1-\delta \\ 
v\Gamma _{k} & 0 & 1-\delta & V%
\end{array}%
\right) ,
\end{equation}%
and%
\begin{equation}
B=w\left( 
\begin{array}{cccc}
0 & \delta -1 & 0 & \Gamma _{-k}/w \\ 
1-\delta & 0 & -\Gamma _{0}/w & 0 \\ 
0 & \Gamma _{0}/w & 0 & \delta -1 \\ 
-\Gamma _{k}/w & 0 & 1-\delta & 0%
\end{array}%
\right) ,
\end{equation}%
where the $k$-dependent factor $\Gamma _{k}=\left( 1+\delta \right) e^{ik}$.
The symmetry of the matrix $h_{k}$ contains the core information of the
system, which allows us to obtain the conclusion without solving the $%
8\times 8$ matrix $h_{k}$\ explicitly.

Introducing the $8\times 8$\ matrix%
\begin{equation}
P=\left( 
\begin{array}{cc}
0 & I_{2} \\ 
I_{2} & 0%
\end{array}%
\right) ,I_{2}=\left( 
\begin{array}{cc}
1 & 0 \\ 
0 & 1%
\end{array}%
\right) ,
\end{equation}%
we readily have%
\begin{equation}
P\left( 
\begin{array}{cc}
A & -B \\ 
B & -A%
\end{array}%
\right) P^{-1}=-\left( 
\begin{array}{cc}
A & -B \\ 
B & -A%
\end{array}%
\right) ,
\end{equation}%
which ensures that the spectrum $\varepsilon _{k}$ of $h_{k}$\ is symmetric
with respect to the zero-energy point. This can be demonstrated by numerical
simulations for finite-size chains. The spectra of the system with
representative parameters are plotted in {Fig. \ref{fig1} (b2).}
Furthermore, $\left\vert \varepsilon _{k}\right\vert $ can be obtained
directly from the diagonalization of the $4\times 4$ matrix $\left(
A+B\right) \left( A-B\right) $ (details are shown in the Appendix).

In the following, we focus on the degeneracy points of $\varepsilon _{k}$\
in the parameter space. The derivations in the Appendix show that the band
degenerate points\ always lay in the subspace with $k=0$ or $\pi $, i.e.,
the degenerate zero-energy points (nodal line) can be determined by 
\begin{equation}
\varepsilon _{0}\varepsilon _{\pi }=0.
\end{equation}%
Then, the corresponding nodal lines obey the equations

\begin{equation}
\left( \frac{\delta +c_{1}}{a_{1}}\right) ^{2}+\left( \frac{V}{b_{1}}\right)
^{2}=1,  \label{ellipse}
\end{equation}%
and%
\begin{equation}
\left( \frac{\delta +c_{2}}{a_{2}}\right) ^{2}-\left( \frac{V}{b_{2}}\right)
^{2}=1,
\end{equation}%
which are obviously an ellipse and a hyperbola, respectively. Here, the
shapes of two conic curves are determined by the system parameters%
\begin{eqnarray}
a_{1} &=&\frac{2\left( w+v\right) }{4-\left( w-v\right) ^{2}},a_{2}=\frac{2}{%
w+v}, \\
b_{1} &=&\frac{2\left( w+v\right) }{\sqrt{4-\left( w-v\right) ^{2}}},b_{2}=2,
\\
c_{1} &=&\frac{w^{2}-v^{2}}{4-\left( w-v\right) ^{2}},c_{2}=\frac{v-w}{v+w}.
\end{eqnarray}%
We note that when taking $w+v\rightarrow 0$ (see also the exact solution for 
$w+v=0$ in the following section), the first equation reduces to a point $%
\left( \delta, V \right) =\left( 0,0\right) $ if $\left\vert w\right\vert <1$%
. In this work, we focus on the nodal line around the point $\left(\delta,V
\right) =\left( 0,0\right) $. The corresponding nodal lines and energy band
edges are plotted in Fig. \ref{fig2} (a) and (b). In the following, we will
investigate the topological features related to the nodal lines. This study
differs from all the previous work on nodal lines, which are lines in 3D
space, while the present study appears only in 2D space.

\begin{figure*}[tbh]
\centering \includegraphics[width=0.9\textwidth]{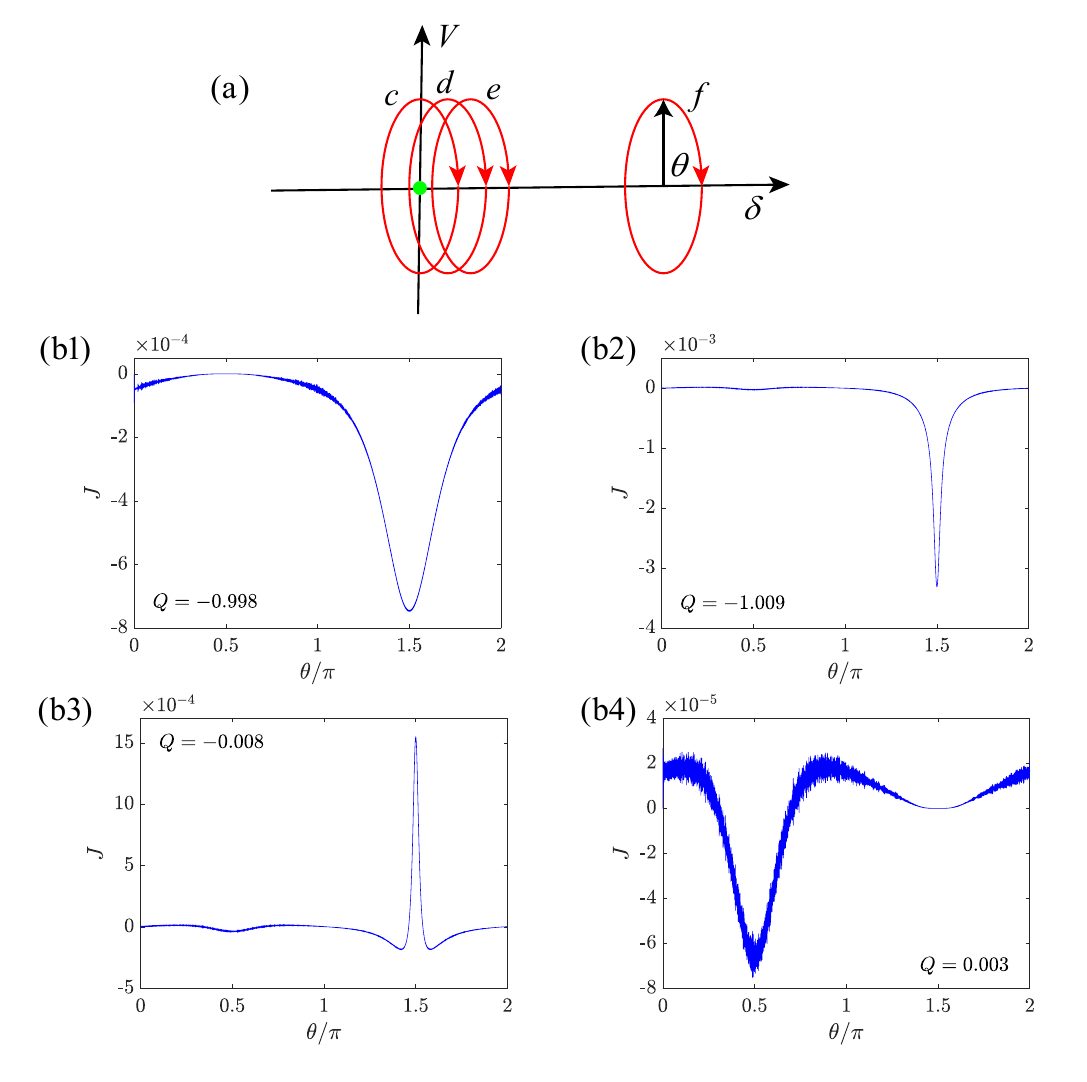}
\caption{The adiabatic current in Eq. (\protect\ref{J}) and pumping charge
in Eq. (\protect\ref{Q}) of the ground state of the Hamiltonian with a
single degenerate point for different passages. (a) The green solid dot is
the degenerate point $\left( \protect\delta ,V\right) =\left( 0,0\right) $
and the red lines represent different adiabatic loops. The equations are $%
\protect\delta =\sin \left( \protect\omega t\right)+\protect\delta_{0}$, $%
V=3\cos \left( \protect\omega t\right) $ and $\protect\theta =\protect\omega %
t$; for $c, d, e,$ and $f $, the parameters $\protect\delta_{0}$ are 0, 0.9,
1.1, and 2, respectively; and (b1), (b2), (b3) and (b4) show the
corresponding numerical results of the adiabatic current and pumping charge.
The time interval is $t_{l}-t_{l-1}=0.0628$, and the other parameters are $%
w=0.6,v=-0.6$, $\protect\omega= 0.001$, and $N=50$. This implies that if the
adiabatic loop encircles the degenerate point, the pumping charge is nearly
-1; otherwise, the pumping charge is nearly 0.}
\label{fig3}
\end{figure*}

\section{Equivanlent RM model}

\label{Equivanlent RM model}

In this section, we start with a Kitaev model that connects to a well-known
model possessing topological features. We consider the Hamiltonian $H$ with $%
v=-w$, at which the Hamiltonian becomes a simplified form%
\begin{equation}
h_{k}=h_{1}^{k}+h_{2}^{k}=\sum_{l=1,2}\left( 
\begin{array}{cc}
A_{l} & -B_{l} \\ 
B_{l} & -A_{l}%
\end{array}%
\right) ,
\end{equation}%
and obey the commutative relations

\begin{equation}
\left[ h_{1}^{k},h_{2}^{k}\right] =\left[ H_{1},H_{2}\right] =0.
\end{equation}%
The relations can be checked by a straightforward derivation for the
following explicit forms of four matrices. Here, two $k$-independent $%
4\times 4$ matrices are

\begin{equation}
A_{1}=\left( 
\begin{array}{cccc}
V & 1-\delta & 0 & 0 \\ 
1-\delta & -V & 0 & 0 \\ 
0 & 0 & -V & 1-\delta \\ 
0 & 0 & 1-\delta & V%
\end{array}%
\right) ,
\end{equation}%
and%
\begin{equation}
B_{2}=w\left( 1-\delta \right) \left( 
\begin{array}{cccc}
0 & -1 & 0 & 0 \\ 
1 & 0 & 0 & 0 \\ 
0 & 0 & 0 & -1 \\ 
0 & 0 & 1 & 0%
\end{array}%
\right) ,
\end{equation}%
while the other two $k$-dependent ones are%
\begin{equation}
B_{1}=\left( 1+\delta \right) \left( 
\begin{array}{cccc}
0 & 0 & 0 & e^{-ik} \\ 
0 & 0 & -1 & 0 \\ 
0 & 1 & 0 & 0 \\ 
-e^{ik} & 0 & 0 & 0%
\end{array}%
\right) ,
\end{equation}%
and%
\begin{equation}
A_{2}=-\left( 1+\delta \right) w\left( 
\begin{array}{cccc}
0 & 0 & 0 & e^{-ik} \\ 
0 & 0 & 1 & 0 \\ 
0 & 1 & 0 & 0 \\ 
e^{ik} & 0 & 0 & 0%
\end{array}%
\right) .
\end{equation}%
Obviously, $H$ and $H_{1}$ have common eigenstates. Importantly, the
derivation in the Appendix shows that $H$ and $H_{1}$ have the same ground
state within the region $\left\vert w\right\vert <1$. This means that one
can focus on the investigation and analysis of $H_{1}$ only. Notably, we
will show that $H_{1}$ has a connection to a RM model.

Taking a partial particle-hole transformation

\begin{equation}
\left( 
\begin{array}{c}
c_{4j-3} \\ 
c_{4j-2} \\ 
c_{4j-1} \\ 
c_{4j}%
\end{array}%
\right) \longrightarrow \left( 
\begin{array}{c}
c_{4j-3}^{\dag } \\ 
c_{4j-2}^{\dag } \\ 
c_{4j-1} \\ 
-c_{4j}%
\end{array}%
\right) ,  \label{p-h Trans.}
\end{equation}%
with $j\in \left[ 1,N\right] $, $H_{1}$ becomes a RM model \cite{Rice}%
\begin{eqnarray}
H_{\mathrm{RM}} &=&-\sum_{l=1}^{2N}\left[ \left( 1-\delta \right)
c_{2l-1}^{\dag }c_{2l}+\left( 1+\delta \right) c_{2l}^{\dag }c_{2l+1}\right]
+\mathrm{h.c.}  \notag \\
&&+V\sum_{l=1}^{2N}\left( -c_{2l-1}^{\dag }c_{2l-1}+c_{2l}^{\dag
}c_{2l}\right) .
\end{eqnarray}%
It is well-known that $H_{\mathrm{RM}}$ is one of the basic models discussed
in connection with ferroelectrics \cite{smith1993theory,onoda2004hall} and
topological properties. This provides a natural platform for directly
studying topological invariants through dynamics both from theoretical \cite%
{XiaoDi,WR1,WR2} and experimental \cite{Marcos} perspectives. For a RM
model, it has been shown \cite{XiaoDi} that if the system adiabatically
evolves along a loop enclosing the degeneracy points $(0,0)$ in the $\delta
-V$ plane, then the polarization will change by $\pm 1$, where the sign
depends on the direction of the loop. On the other hand, if the loop does
not contain the degeneracy point, then the pumped charge is zero \cite%
{XiaoDi}.

Specifically, considering a time-dependent RM Hamiltonian $H_{\mathrm{RM}%
}(t) $ with $\delta (t)=\delta (T+t)$\ and $V(t)=V(T+t)$, the pumping charge
passing site $j$ for the evolved state $\left\vert \phi (t)\right\rangle $
from the initial ground state $\left\vert \phi (0)\right\rangle $ of $H_{%
\mathrm{RM}}(0)$ for the time evolution period $T$ can be expressed as 
\begin{equation}
Q_{j}=\int_{0}^{T}\langle \phi (t)|\mathcal{J}_{j}\left\vert \phi
(t)\right\rangle \mathrm{d}t,
\end{equation}%
where the current operator is 
\begin{equation}
\mathcal{J}_{j}=-i\left( 1-\delta \right) c_{4j-3}^{\dag }c_{4j-2}+\mathrm{%
h.c.,}
\end{equation}%
where $j=1,2,...,N$. The pumping charge $Q_{j}$ is quantized when it is an
adiabatic cycle, characterizing the topological feature.

Importantly, we note that the current operator $\mathcal{J}_{j}$ is
invariant under the partial particle-hole transformation in Eq. (\ref{p-h
Trans.}). This finding suggested that the dimerized Kitaev chain with $v=-w $
has the same topological features as the RM model within the region $%
\left\vert w\right\vert <1$.

Although this conclusion was obtained rigorously, it is a little surprising
because a RM model supports the conservation of the particle number, while
the Kitaev model does not. To date, almost all studies on the topic of
pumping charge have focused mainly on particle conservation systems. In
addition, the unveiled topology for the special case with $v=-w$ may be
extended to a more general case. This is the main goal of this work.

\begin{figure*}[tbh]
\centering \includegraphics[width=0.9\textwidth]{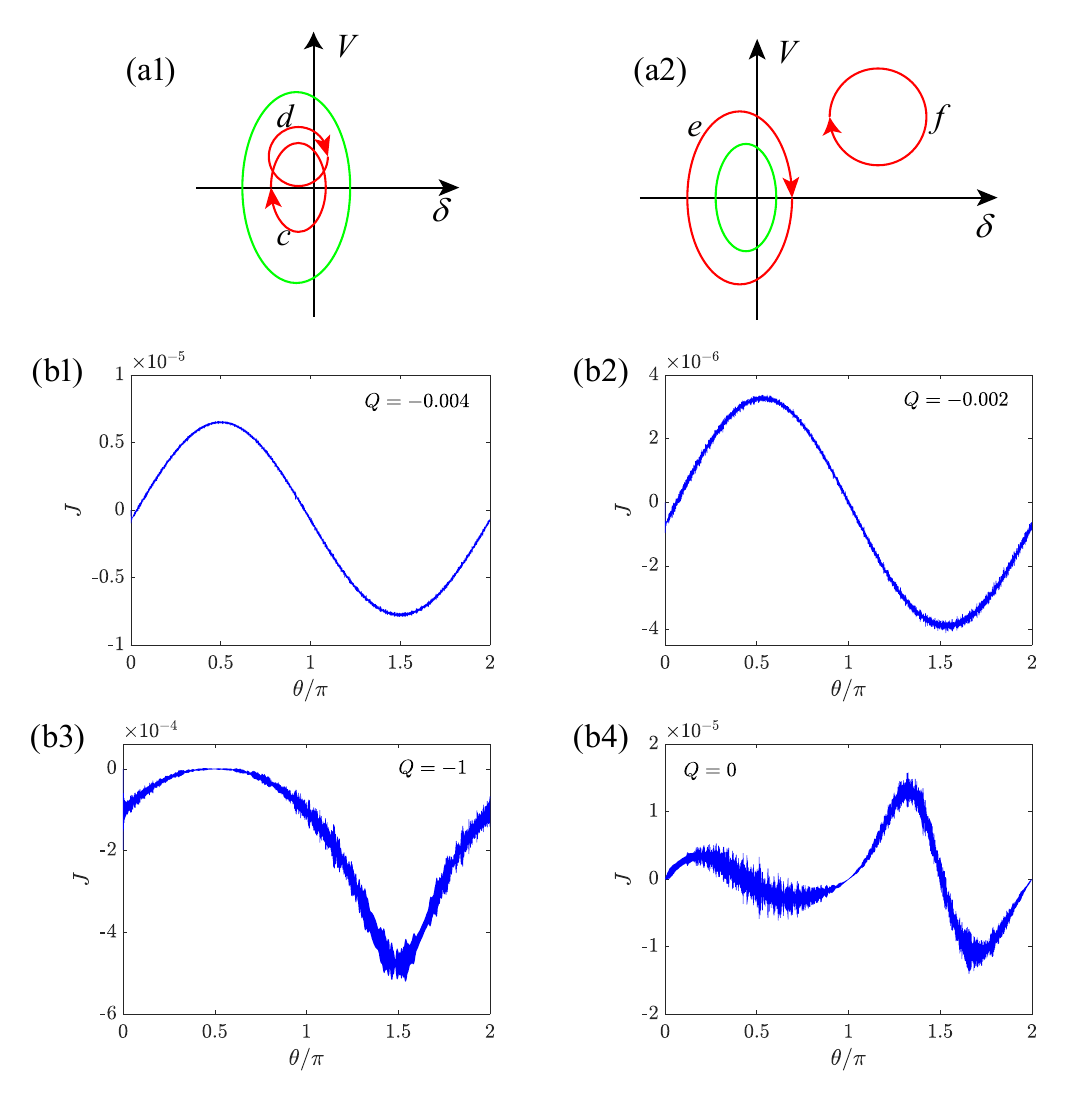}
\caption{The adiabatic current in Eq. (\protect\ref{J}) and pumping charge
in Eq. (\protect\ref{Q}) of the ground state of the Hamiltonian with nodal
lines for different passages. The green loops in (a1) and (a2) are the same
nodal line with the parameters $w=0.6,v=-0.3$ and its center is $\left( 
\protect\delta ,V\right) =\left( \protect\delta _{0},0\right) $. $c, d, e$,
and $f $ are four different adiabatic loops and the corresponding equations
are $c$: $\protect\delta =0.02\sin \left( \protect\omega t\right) +\protect%
\delta _{0}$, $V=0.04\cos \left( \protect\omega t\right) $; $d$: $\protect%
\delta =0.02\sin \left( \protect\omega t\right) +\protect\delta _{0}$, $%
V=0.02\cos \left( \protect\omega t\right) $+0.02; $e$: $\protect\delta =\sin
\left( \protect\omega t\right) +\protect\delta _{0}$, $V=2\cos \left( 
\protect\omega t\right) $ ; $f$: $\protect\delta =0.5\sin \left( \protect%
\omega t\right) +1$, $V=0.5\cos \left( \protect\omega t\right) +1$ and $%
\protect\theta =\protect\omega t$. The other parameters are $N=50$, $\protect%
\omega =0.001$ and the time interval is $t_{l}-t_{l-1}=0.0628$. (b1), (b2),
(b3) and (b4) demonstrate the corresponding numerical results for the
current and pumping charge. This indicates that if the adiabatic loop
encircles the nodal line, the pumping is very close to $-1$, and the pumping
charge is very close to $0$ if the nodal line is outside the adiabatic loop. 
}
\label{fig4}
\end{figure*}

\section{Topology of the nodal ellipse}

\label{Topology of the nodal ellipse} Now, we turn to the question of what
happens in the case with nonzero $v+w$, at which the degeneracy points form
an ellipse. Here, we would like to emphasize that the nodal line lies in a
2D plane in this work rather than in the 3D space in previous works. For a
nodal loop within a 3D space, the pumping charge is studied along a closed
passage piercing the nodal loop. Therefore, we will investigate the topology
associated with the nodal loop from another aspect. Specifically, how does
the pumping charge when an adiabatic loop encloses the ellipse or lies
inside the ellipse? In the latter case, the loop does not enclose any
degeneracy points, there is no doubt that the pumping charge is zero.
However, it is difficult to predict this result for the former case because
the total pumping charge depends on the vortex of an individual degenerate
point \cite{WR1}. In this situation, numerical simulation is an efficient
tool providing evidence for theoretical investigation.

A numerical simulation is performed for the time-dependent Hamiltonian with
the parameters

\begin{equation}
\delta =R_{1}\cos (\omega t)+\delta _{0},V=R_{2}\sin (\omega t)+V_{0},
\end{equation}%
which is an ellipse with a center at ($\delta _{0},V_{0}$) in the $\delta -V$
plane. Here, $\omega $ controls the varying speed of the time-dependent
Hamiltonian. The computation is performed using a uniform mesh in time
discretization. Time $t$ is discretized into $t_{i}$, with $t_{0}=0$ and $%
t_{M}=T$. For a given initial eigenstate $\left\vert \phi \left( 0\right)
\right\rangle $, the time-evolved state is computed using 
\begin{equation}
\left\vert \phi \left( t_{n}\right) \right\rangle =\mathcal{T}%
\prod\limits_{l=1}^{n}\exp [(-iH\left( t_{l-1}\right)
(t_{l}-t_{l-1})]\left\vert \phi (0)\right\rangle ,
\end{equation}%
where $\mathcal{T}$ is the time-order operator. In the simulation, the value
of $M$ is considered sufficiently large to obtain a convergent result. The
total pumping charge passing through site $j$ during the time evolution
period $T$ can be expressed as 
\begin{equation}
Q_{j}\approx \sum_{l=1}^{M}\langle \phi \left( t_{l}\right) |\mathcal{J}%
_{j}\left( t_{l}\right) \left\vert \phi \left( t_{l}\right) \right\rangle
(t_{l}-t_{l-1}).
\end{equation}%
and the corresponding average current and average pumping charge are defined
as 
\begin{equation}
J\left( t_{l}\right) =\frac{1}{N}\sum_{j=1}^{N}\langle \phi \left(
t_{l}\right) |\mathcal{J}_{j}\left( t_{l}\right) \left\vert \phi \left(
t_{l}\right) \right\rangle ,  \label{J}
\end{equation}%
\begin{equation}
Q=\frac{1}{N}\sum_{j=1}^{N}Q_{j}.  \label{Q}
\end{equation}%
Due to the translational symmetry of the system, the current and pumping
charge across the same type of dimer are identical. However, the average
current and pumping charge are favorable for numerical computation.

In principle, the value of $\omega $ should be sufficiently small to fulfill
the requirement of quasiadiabatic evolution. In the computation, we select $%
\omega $ to satisfy the quasiadiabatic condition, under which the obtained
pumping charge is not sensitive to a slight change in $\omega $. Fig. \ref%
{fig3} and Fig. \ref{fig4} present the plots of the simulations for two
kinds of quasiadiabatic passages with the degeneracy point and nodal line
unenclosed and enclosed by a loop, respectively. The instantaneous current
and total accumulated charge of the ground state are computed. According to
our analysis for the case with $w+v=0$, the final value of the pumping
charge $Q$ depends on whether the evolution loop encloses the degeneracy
point $(0,0)$ in the $\delta-V$ parameter space. We first demonstrate this
point in Fig. \ref{fig3}, which clearly shows that $Q=0$ or $-1$ with high
precision for the unenclosed or enclosed circle. We then plot the results
for the case with $w+v\neq 0$ in Fig. \ref{fig4}. Notably, we find that the
pumping charge $Q=0$ or $-1$ is highly precise for unenclosed or enclosed
loops. This strongly implies that the pumping charge can be considered a
quasitopological invariant arising from a nodal ellipse. Thus far, this
finding has not been verified theoretically.

\section{Summary}

\label{summary}

In summary, we investigated the topology associated with nodal loops in a
system without particle number conservation. In addition, the nodal loop
lies in a 2D parameter space rather than 3D space. In previous work on a
nodal line in 3D space, such as that in Ref. \cite{WR1}, the related
topology feature was essentially the same as that of an isolated degenerate
point. This can be seen from a cross section in 3D space, in which the nodal
line reduces to a degenerate point. As an example, we studied the
topological pumping charge of a dimerized Kitaev chain with spatially
modulated chemical potential. This model has the advantage that it can be
mapped into a Rice-Mele model by a partial particle-hole transformation
under certain constraints. This motivates us to compute the pumping charge
beyond this special case. Numerical simulation of a finite-size chain
indicates that the pumping charge is zero for a quasiadiabatic loop within
the nodal loop and $\pm 1 $ for the passage loop enclosing the nodal loop.
This indicates that such a Kitaev model supports topological charge pumping
as a demonstration of Chern number. Our findings unveil a hidden topology in
a class of Kitaev chains, exploring topological matter from an alternative
aspect.

\section*{Acknowledgements}

This work was supported by the National Natural Science Foundation of China
(under Grant No. 12374461).

\appendix

\section{Appendix}

In this appendix, we derive the spectrum of the Hamiltonian and show that
(i) $H$ and $H_{1}$ with $v=-w$ have the same ground state within the region 
$\left\vert w\right\vert <1$ and that (ii) the degenerate lines are a set of
conic functions.

(i) In the case of $v+w=0$, we have $\left[ H,H_{1}\right] =\left[
H_{1},H_{2}\right] =0$. The corresponding core matrices of $H_{1}$\ and $%
H_{2}$ are%
\begin{equation}
h_{l}^{k}=\left( 
\begin{array}{cc}
A_{l} & -B_{l} \\ 
B_{l} & -A_{l}%
\end{array}%
\right) ,l=1,2.
\end{equation}%
Which are defined in the main text. We note that the spectrum $\left\{
\varepsilon _{l}^{k},l=1,2\right\} $ of $h_{l}^{k}$ is also symmetric with
respect to the zero-energy point; then, one can obtain $\left\vert
\varepsilon _{l}^{k}\right\vert $ by solving the matrix%
\begin{equation}
\left( h_{l}^{k}\right) ^{2}=\left( 
\begin{array}{cc}
A_{l}^{2}-B_{l}^{2} & B_{l}A_{l}-A_{l}B_{l} \\ 
B_{l}A_{l}-A_{l}B_{l} & A_{l}^{2}-B_{l}^{2}%
\end{array}%
\right) .
\end{equation}

Taking the unitary transformation%
\begin{equation}
u=\frac{1}{\sqrt{2}}\left( 
\begin{array}{cc}
1 & 1 \\ 
1 & -1%
\end{array}%
\right) ,
\end{equation}%
we have%
\begin{equation}
u\left( h_{l}^{k}\right) ^{2}u^{\dag }=\left( 
\begin{array}{cc}
D_{l}^{+}D_{l}^{-} & 0 \\ 
0 & D_{l}^{-}D_{l}^{+}%
\end{array}%
\right) ,
\end{equation}%
where the diagonal block matrices are 
\begin{equation}
D_{l}^{\pm }=A_{l}\pm B_{l}.
\end{equation}%
Solving the equation 
\begin{equation}
\det \left[ D_{l}^{+}D_{l}^{-}-\left( \varepsilon _{l}^{k}\right) ^{2}I_{2}%
\right] =0,  \label{det}
\end{equation}%
we obtain 
\begin{eqnarray}
\left( \varepsilon _{1}^{k}\right) ^{2} &=&V^{2}+\left( \varepsilon
_{2}^{k}\right) ^{2}/w^{2}, \\
\left( \varepsilon _{2}^{k}\right) ^{2} &=&w^{2}\left[ 2\left( \delta
^{2}+1\right) \pm f\left( \delta ,k\right) \right] .
\end{eqnarray}%
with the function $f\left( \delta ,k\right) =\left( \delta ^{2}-1\right) 
\sqrt{2\left( 1+\cos k\right) }$. Obviously, we always have $\left\vert
\varepsilon _{1}^{k}\right\vert >\left\vert \varepsilon _{2}^{k}\right\vert $
when $\left\vert w\right\vert <1$. As a result, the ground state of $H_{1}$
is also the ground state of $H$.

(ii) For the general case with arbitrary $w$ and $v$, we still have%
\begin{equation}
\det \left[ D^{+}D^{-}-\left( \varepsilon _{k}\right) ^{2}I_{2}\right] =0,
\end{equation}%
based on a similar analysis. Then, the degeneracy line can be determined by 
\begin{equation}
\det \left( D^{+}D^{-}\right) =0.
\end{equation}%
Here, matrices $D^{\pm }=A\pm B$ can be written in the explicit form%
\begin{equation}
D^{\pm }=\left( 
\begin{array}{cccc}
V & \Sigma ^{\mp } & 0 & \Delta ^{\pm }e^{-ik} \\ 
\Sigma ^{\pm } & -V & \Delta ^{\mp } & 0 \\ 
0 & \Delta ^{\pm } & -V & \Sigma ^{\mp } \\ 
\Delta ^{\mp }e^{ik} & 0 & \Sigma ^{\pm } & V%
\end{array}%
\right) ,
\end{equation}%
where $\Sigma ^{\pm }=\left( 1-\delta \right) \left( 1\pm w\right) $ and $%
\Delta ^{\pm }=\left( 1+\delta \right) \left( v\pm 1\right) $. The equation $%
\det \left( D^{\pm }\right) =0$ requires the vanishing real and imaginary
parts of $\det \left( D^{\pm }\right) $, or explicitly%
\begin{eqnarray}
\left( V^{2}+\Sigma ^{+}\Sigma ^{-}-\Delta ^{+}\Delta ^{-}\right)
^{2}+2\Sigma ^{+}\Sigma ^{-}\Delta ^{+}\Delta ^{-} &&  \notag \\
-\left[ \left( \Sigma ^{+}\Delta ^{+}\right) ^{2}+\left( \Sigma ^{-}\Delta
^{-}\right) ^{2}\right] \cos k_{c}=0, &&
\end{eqnarray}%
and%
\begin{equation}
\left[ \left( \Sigma ^{+}\Delta ^{+}\right) ^{2}-\left( \Sigma ^{-}\Delta
^{-}\right) ^{2}\right] \sin k_{c}=0.
\end{equation}%
Then, the degeneracy lines obey the conditions%
\begin{equation}
\left( V^{2}+\Sigma ^{+}\Sigma ^{-}-\Delta ^{+}\Delta ^{-}\right) ^{2}=\mp
\left( \Sigma ^{+}\Delta ^{+}\pm \Sigma ^{-}\Delta ^{-}\right) ^{2},
\end{equation}%
for $k_{c}=\pi /2\pm \pi /2$. In the following, we consider only the cases
with $\left\vert w\right\vert <1$ and $\left\vert v\right\vert <1$ for
simplicity and without loss of practicality. Within this region, we always
have $V^{2}+\Sigma ^{+}\Sigma ^{-}-\Delta ^{+}\Delta ^{-}>0$. Then, the
conditions for degeneracy lines become%
\begin{equation}
\left\vert V^{2}+\Sigma ^{+}\Sigma ^{-}-\Delta ^{+}\Delta ^{-}\right\vert
=\left\vert \Sigma ^{+}\Delta ^{+}-\Sigma ^{-}\Delta ^{-}\right\vert ,
\end{equation}%
which are simplified to two conic functions 
\begin{equation}
\left( \frac{\delta +c_{1}}{a_{1}}\right) ^{2}+\left( \frac{V}{b_{1}}\right)
^{2}=1,
\end{equation}%
and 
\begin{equation}
\left( \frac{\delta +c_{2}}{a_{2}}\right) ^{2}-\left( \frac{V}{b_{2}}\right)
^{2}=1,
\end{equation}%
where the parameters are 
\begin{eqnarray}
a_{1} &=&\frac{2\left( w+v\right) }{4-\left( w-v\right) ^{2}},a_{2}=\frac{2}{%
w+v}, \\
b_{1} &=&\frac{2\left( w+v\right) }{\sqrt{4-\left( w-v\right) ^{2}}},b_{2}=2,
\\
c_{1} &=&\frac{w^{2}-v^{2}}{4-\left( w-v\right) ^{2}},c_{2}=\frac{v-w}{v+w}.
\end{eqnarray}%
In conclusion, the nodal lines are conic functions associated with the
zero-momentum invariant subspace.

\end{document}